\documentclass[%
reprint,
amsmath,
amssymb,
aps,
]{revtex4-1}

\usepackage{graphicx}
\usepackage{dcolumn}
\usepackage{bm}

\begin{document}
\preprint{APS/123-QED}

\title{\boldmath Determination of the number of $J/\psi$ events
with $J/\psi \rightarrow inclusive$ decays}

\author{
 M.~Ablikim$^{1}$, M.~N.~Achasov$^{5}$, D.~J.~Ambrose$^{40}$, F.~F.~An$^{1}$, Q.~An$^{41}$, Z.~H.~An$^{1}$,
J.~Z.~Bai$^{1}$, Y.~Ban$^{27}$, J.~Becker$^{2}$, N.~Berger$^{1}$
% ...
}
\affiliation{
$^{1}$ Institute of High Energy Physics, Beijing 100049, China\\
$^{2}$ Bochum Ruhr-University, 44780 Bochum, Germany\\
% ...
$^{5}$ G.I. Budker Institute of Nuclear Physics SB RAS (BINP), Novosibirsk 630090, Russia\\
% ...
$^{27}$ Peking University, Beijing 100871, China\\
% ...
$^{40}$ University of Rochester, Rochester, New York 14627, USA\\
$^{41}$ University of Science and Technology of China, Hefei 230026, China\\
% ...
}

\collaboration{BESIII Collaboration}

\begin{abstract}
The number of $J/\psi$ events collected with the BESIII detector at
the BEPCII from June 12 to July 28, 2009 is determined to be
$(225.3\pm2.8)\times10^{6}$ using $J/\psi \rightarrow inclusive$
events, where the uncertainty is the systematic error and the
statistical one is negligible.
\end{abstract}


\maketitle
\section{Introduction}

To meet the challenge of precision measurements of $\tau-$charm
physics, a major upgrade on the Beijing Electron-Positron Collider
(BEPC) and the Beijing Spectrometer (BES) was completed in 2008 (now
called BEPCII and BESIII). BEPCII is a double ring $e^+e^-$ collider
with a design peak luminosity of $10^{33}$ cm$^{-2}s^{-1}$ at $\sqrt s = $3.773
GeV, which is 100 times that of its predecessor. The BESIII detector
is a large solid-angle magnetic spectrometer that is described in
detail in Ref.~\cite{bes3}. The major improvements in the BES detector
are the huge superconducting solenoid magnet with a central field of 1
T, which offers a significant improvement in the momentum resolution
of charged particles, and a cesium iodide (CsI) calorimeter for the energy
measurement of electrons and photons, which provides more than a
factor of 10 improvement in the precision of electromagnetic shower
energy measurements.

Since the discovery of the $J/\psi$ in 1974, it has always been
regarded as an ideal laboratory to study light hadron spectroscopy and
to search for new types of hadrons (e.g. glueballs, hybrids and
exotics). With 58 million $J/\psi$ events collected with the BESII
detector, many important results have been obtained, which underlined
the importance of the study of $J/\psi$ decays. Therefore, after a
successful commissioning of the BESIII detector together with BEPCII,
a large sample of $J/\psi$ events was collected from June 12 to July
28, 2009, which allows the study of the properties and the decays of
the $J/\psi$ with unprecedented precision.

The number of $J/\psi$ events and its uncertainty are two key
quantities in the precision measurements of $J/\psi$ decays. At BESII,
the number of $J/\psi$ events was determined with $J/\psi\rightarrow$
4-prong events, and its systematic uncertainty was
4.7\%~\cite{bes2tot}. The excellent BESIII detector and its good
performance allow the determination of the number of $J/\psi$ events
with higher precision. To reduce the systematic uncertainty
from that in Ref.~\cite{bes2tot}, a new method using $J/\psi\rightarrow
inclusive$ events is introduced.
The number of $J/\psi$ events ($N_{J/\psi}$) is calculated
with
\begin{eqnarray}
\label{Nojpsi}
N_{J/\psi}=\frac{N_{sel}-N_{bg}}{\epsilon_{trig} \times
\epsilon^{\psi^{\prime}}_{data} \times f_{cor}},
\end{eqnarray}
where $N_{sel}$ is the number of $J/\psi \rightarrow inclusive$
events selected from $J/\psi$ data; $N_{bg}$ is the number of
background events estimated from the continuum data taken at the
center-of-mass energy of 3.08 GeV; $\epsilon_{trig}$ is the trigger
efficiency; $\epsilon^{\psi^{\prime}}_{data}$ is the $J/\psi
\rightarrow inclusive$ detection efficiency determined experimentally
from $\psi^{\prime}$ data using $\psi^{\prime}\rightarrow \pi^+ \pi^-
J/\psi$ events; $f_{cor}$ is a correction factor for
$\epsilon^{\psi^{\prime}}_{data}$, obtained from Monte Carlo (MC)
simulation which accounts for the difference between the $J/\psi$
events produced at rest and those produced from
$\psi^{\prime}\rightarrow \pi^+ \pi^- J/\psi$.  The correction factor
in Eq.~(\ref{Nojpsi}), which is approximately unity, is determined
from
\begin{eqnarray}
\label{Fcor}
f_{cor} = \frac {\epsilon^{J/\psi}_{mc}}  {\epsilon^{\psi^{\prime}}_{mc}},
\end{eqnarray}
where $\epsilon^{J/\psi}_{mc}$ is the detection efficiency of
$J/\psi\rightarrow inclusive$ events determined from the $J/\psi$ MC
sample and $\epsilon^{\psi^{\prime}}_{mc}$ is the efficiency
determined from the $\psi^{\prime}\to \pi^+\pi^- J/\psi ~(J/\psi
\rightarrow inclusive)$ MC sample.

There are two major improvements over the method in
Ref.~\cite{bes2tot}. One is the generalization of the
$J/\psi\rightarrow$ 4-prong events to $J/\psi\rightarrow inclusive$
events, which allows the number of $J/\psi$ events to be determined
by requiring different numbers of charged tracks; the other is to use the
MC samples of $J/\psi\rightarrow inclusive$ and
$\psi^{\prime}\rightarrow \pi^+ \pi^- J/\psi ~(J/\psi\rightarrow
inclusive)$ events generated with the BesEvtGen
generator~\cite{evtgen} based on GEANT4~\cite{geant4} to determine the
correction factor, $f_{cor}$. In this analysis, the events with more than
one charged tracks are used to determine the number of $J/\psi$ events.

At present only about 50\% of the $J/\psi$ decays are observed and
listed in the Particle Data Group tables (PDG)~\cite{PDG}. In the MC
simulation package, the unknown $J/\psi$ decays are roughly generated
with the Lundcharm model. In the Lundcharm model, charmonium decay via
gluons is described by the QCD partonic theory, and the partonic
hadronization is handled by the LUND model. Extended $C$- and
$G$-parity conservation are assumed and abnormal suppression effects
of charmonium decay are included~\cite{LUND}.

\section{\label{hadsel}  $J/\psi\rightarrow inclusive$ selection criteria}

Event selection criteria are required to
distinguish $J/\psi\rightarrow inclusive$ events from
 Bhabha ($e^+e^-\rightarrow e^+e^-$), dimuon ($e^+e^-\rightarrow\mu^+\mu^-$),
cosmic ray and beam-gas events in $J/\psi$ data.

At the track level, candidate events are required to satisfy the
following selection criteria:
\begin{enumerate}
\item
 Charged tracks are reconstructed using hits in the Main Drift Chamber
(MDC) and are required to be in the polar angle range $|\cos\theta| <
0.93$, have momentum $p<2.0$ GeV/$c$, and have the point of closest
approach of the track to the beamline within 15 cm of the
interaction point along the beam direction ($V_{z}$) and within 1 cm
in the plane perpendicular to the beam ($V_{r}$).
\item Clusters in the electromagnetic calorimeter (EMC) must have
at least 25 (50) MeV of energy in the barrel (end cap) EMC, have
$|\cos\theta| < 0.83$ in the barrel ($0.86< |\cos\theta| < 0.93$ in the endcap),
and have EMC cluster timing $T$ in the range
of $0<T<15$ (with unit of 50 ns) to suppress electronic noise and energy deposits unrelated
to the event.
\end{enumerate}

\begin{figure}[b]
\includegraphics[width=7.0cm,height=5.0cm]{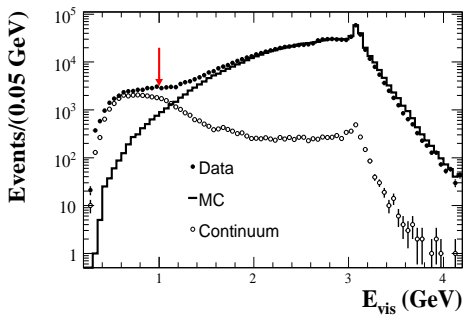}
\caption{\label{evis} The visible energy distributions for $J/\psi$
data (dots with error bars), continuum data (circles with error bars)
and MC simulation of $J/\psi \rightarrow inclusive$ events
(histogram). The arrow indicates the minimum $E_{vis}$ required to
select inclusive events.  }
\end{figure}

At the event level, at least two charged tracks are required, and the
visible energy, $E_{vis}$, must be greater than 1.0 GeV. Here
$E_{vis}$ is defined as the sum of charged particle energies computed
from the track momenta by assuming pion masses and the neutral shower
energies deposited in the EMC. According to the distribution of
visible energy shown in Fig.~\ref{evis}, this requirement removes two
thirds of background events, estimated with the continuum data taken
at the center-of-mass energy of 3.08 GeV, while it has little effect
on the inclusive events.

\begin{figure}[b]
\includegraphics[width=7.0cm,height=5.0cm]{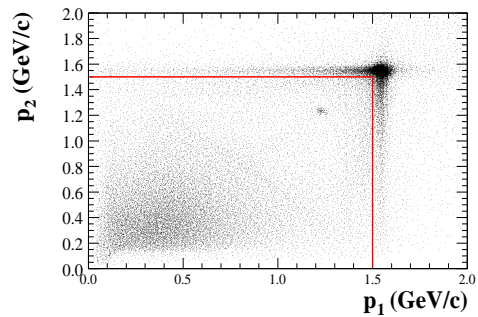}
\caption{\label{pcutn} The scatter plot of the momenta of the
charged tracks for 2-prong events. The cluster around 1.55 GeV/$c$
corresponds to the contribution from leptonic pairs.  Most are removed
with the requirements on the two charged tracks, $p_1<1.5$ GeV/$c$ and
$p_2<1.5$ GeV/$c$, as indicated by the solid lines.}
\end{figure}

\begin{figure}
\includegraphics[width=7.0cm,height=5.0cm]{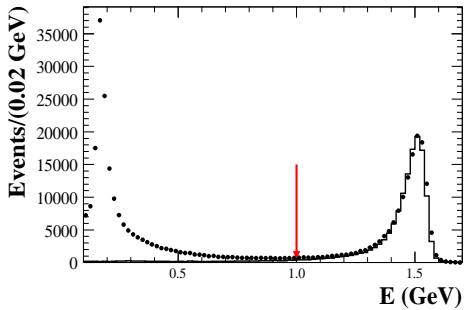}
\caption{\label{etrk} The distributions of deposited energy in the
  EMC by the charged tracks of 2-prong events for $J/\psi$ data (dots
  with error bars) and for the combined, normalized MC simulations of
  $e^+e^-\rightarrow e^+e^-$ and $J/\psi\rightarrow e^+e^-$
  (histogram).}
\end{figure}

To remove background from Bhabhas and dimuons, events with only two
charged tracks must have the momenta of both charged tracks less than
1.5 GeV/$c$. Fig.~\ref{pcutn} displays the scatter plot of the
momenta of two charged tracks, where the clear cluster with the
momenta around 1.55 GeV/$c$ corresponds to the contribution from leptonic
pairs. Most of the leptonic pairs are removed by the above requirement
as indicated by the solid lines in Fig.~\ref{pcutn}. From the
deposited energy distribution of charged tracks in the EMC, shown in
Fig.~\ref{etrk}, a peak around 1.5 GeV is clearly observed, which
corresponds to the contribution of Bhabha events.  Therefore, to
further remove Bhabha events, the deposited energy in the EMC of each
charged track is required to be less than 1 GeV.
After the momentum and energy selections there remain $174.28 \pm 0.01$
million events ($N_{sel}$) from the $J/\psi$ data. The distributions of
the track parameters for closest approach and track angle $V_r$, $V_z$,
$\cos\theta$,

\begin{figure}[b]
\includegraphics[width=7.0cm,height=5.0cm]{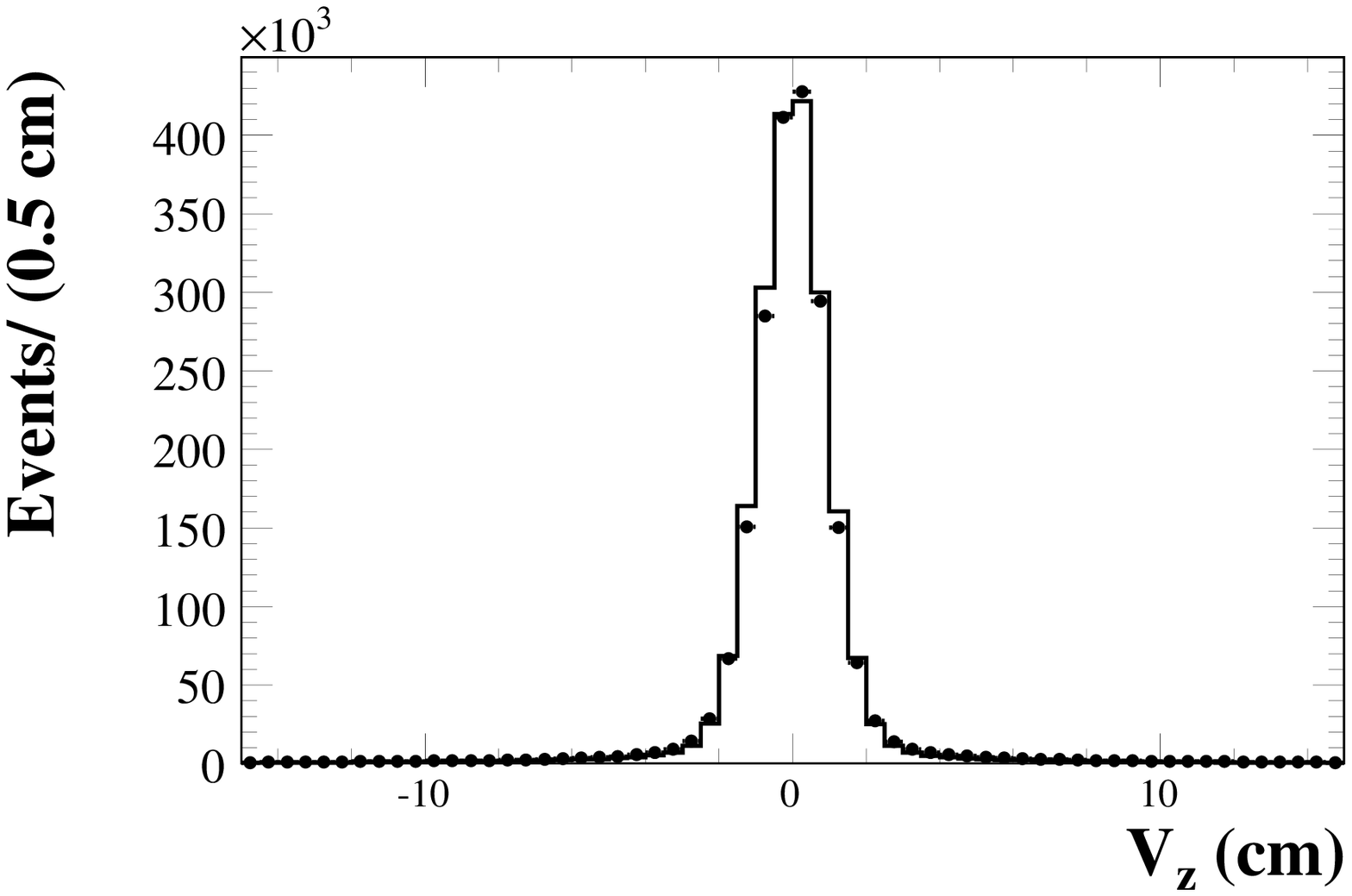}
\caption{\label{vz0} The distributions of $V_z$ for $J/\psi$ data
(dots with error bars) and MC simulation of $J/\psi\rightarrow
inclusive$ (histogram). }
\end{figure}
\begin{figure}[b]
\includegraphics[width=7.0cm,height=5.0cm]{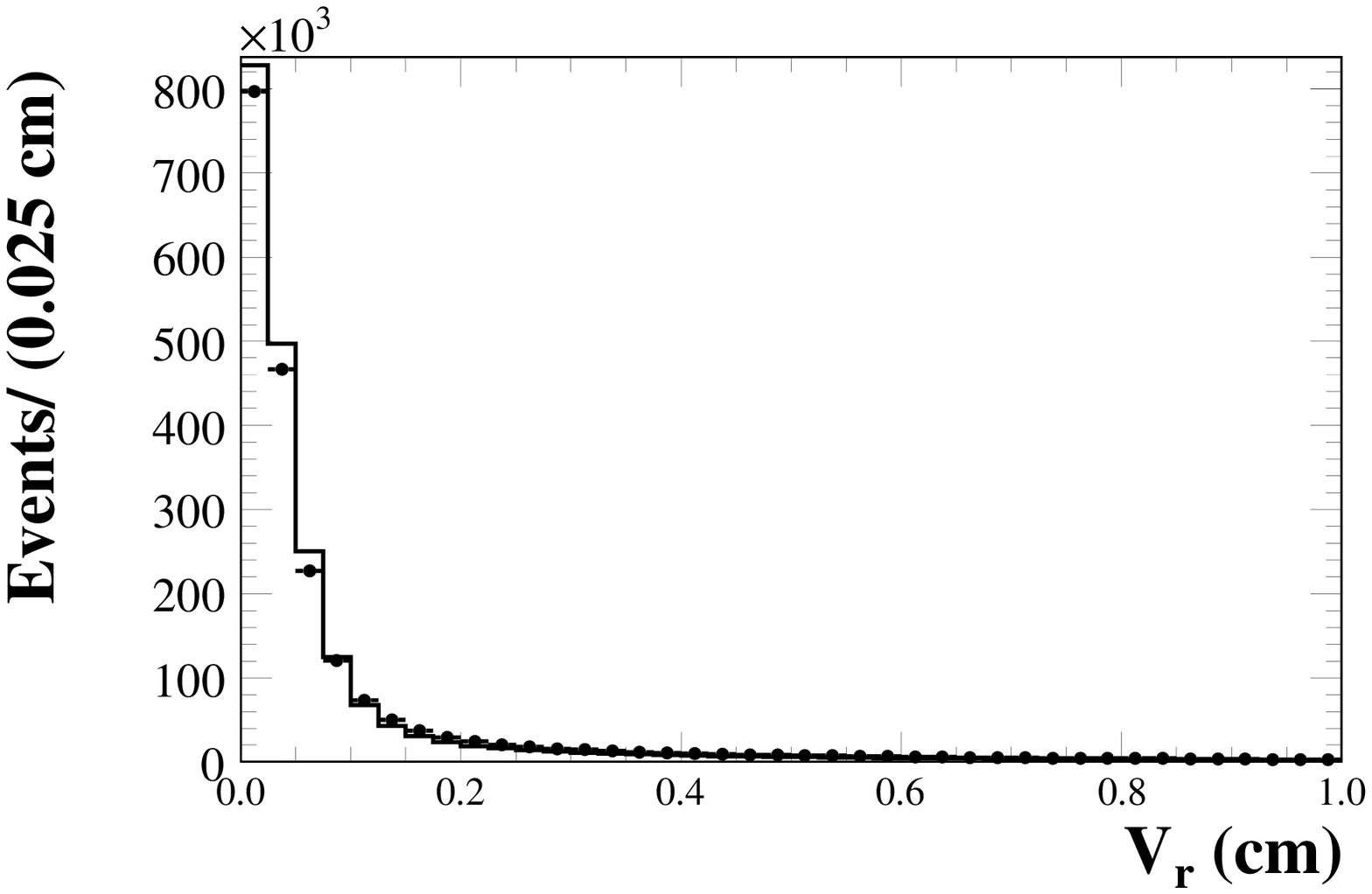}
\caption{\label{vr0} The distributions of $V_r$ for $J/\psi$ data
(dots with error bars) and MC simulation of $J/\psi\rightarrow
inclusive$ (histogram). }
\end{figure}
\begin{figure}[b]
\includegraphics[width=7.0cm,height=5.0cm]{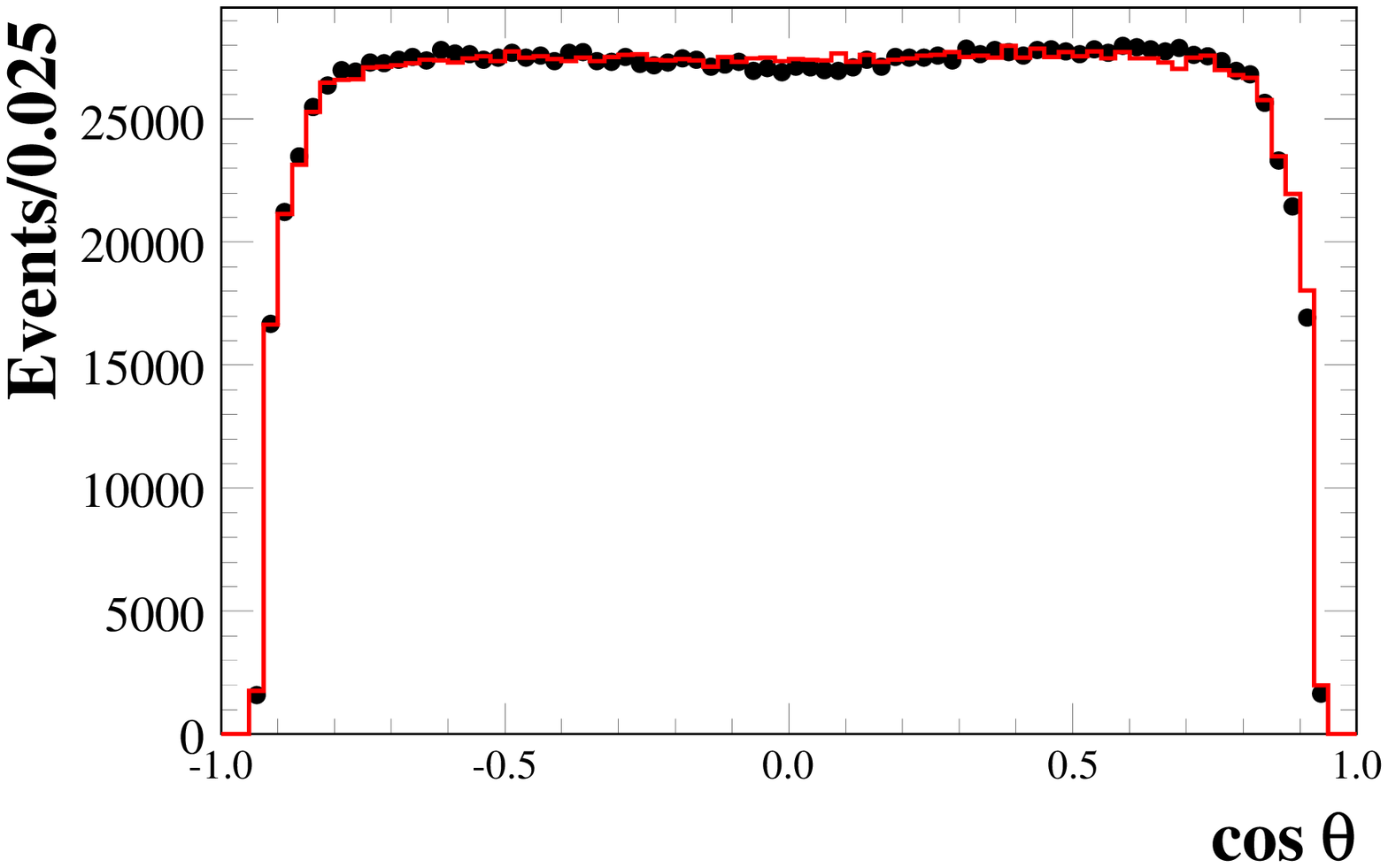}
\caption{\label{cos}The $\cos\theta$ distributions of charged
tracks for $J/\psi$ data (dots with error bars) and MC simulation of
$J/\psi\rightarrow inclusive$ (histogram).}
\end{figure}
\begin{figure}[b]
\includegraphics[width=7.0cm,height=5.0cm]{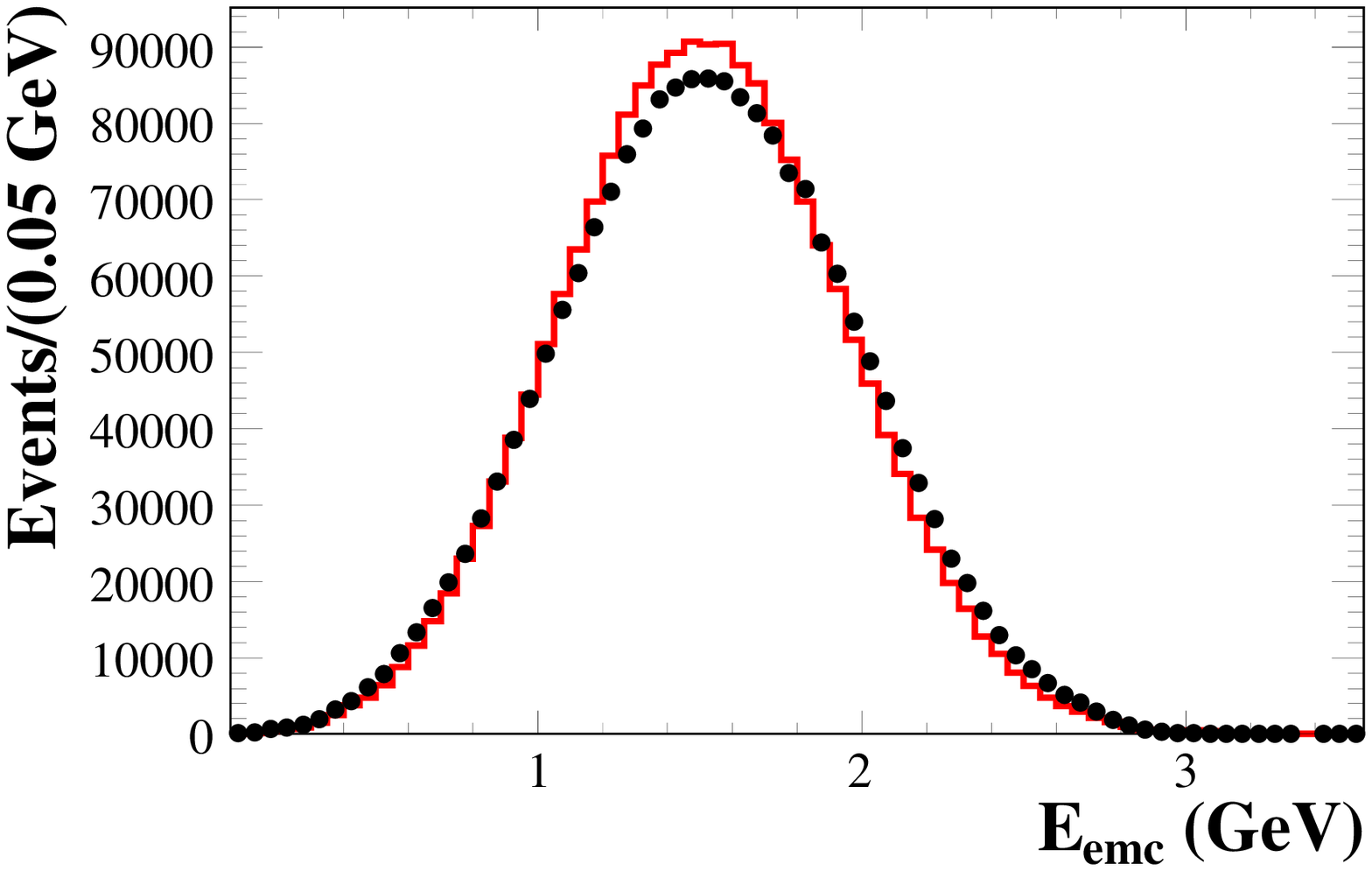}
\caption{\label{emc}The distributions of the total energy
deposited in the EMC of $J/\psi\rightarrow inclusive$ events for
$J/\psi$ data (dots with error bars) and MC simulation (histogram).}
\end{figure}
\begin{figure}[b]
\includegraphics[width=7.0cm,height=5.0cm]{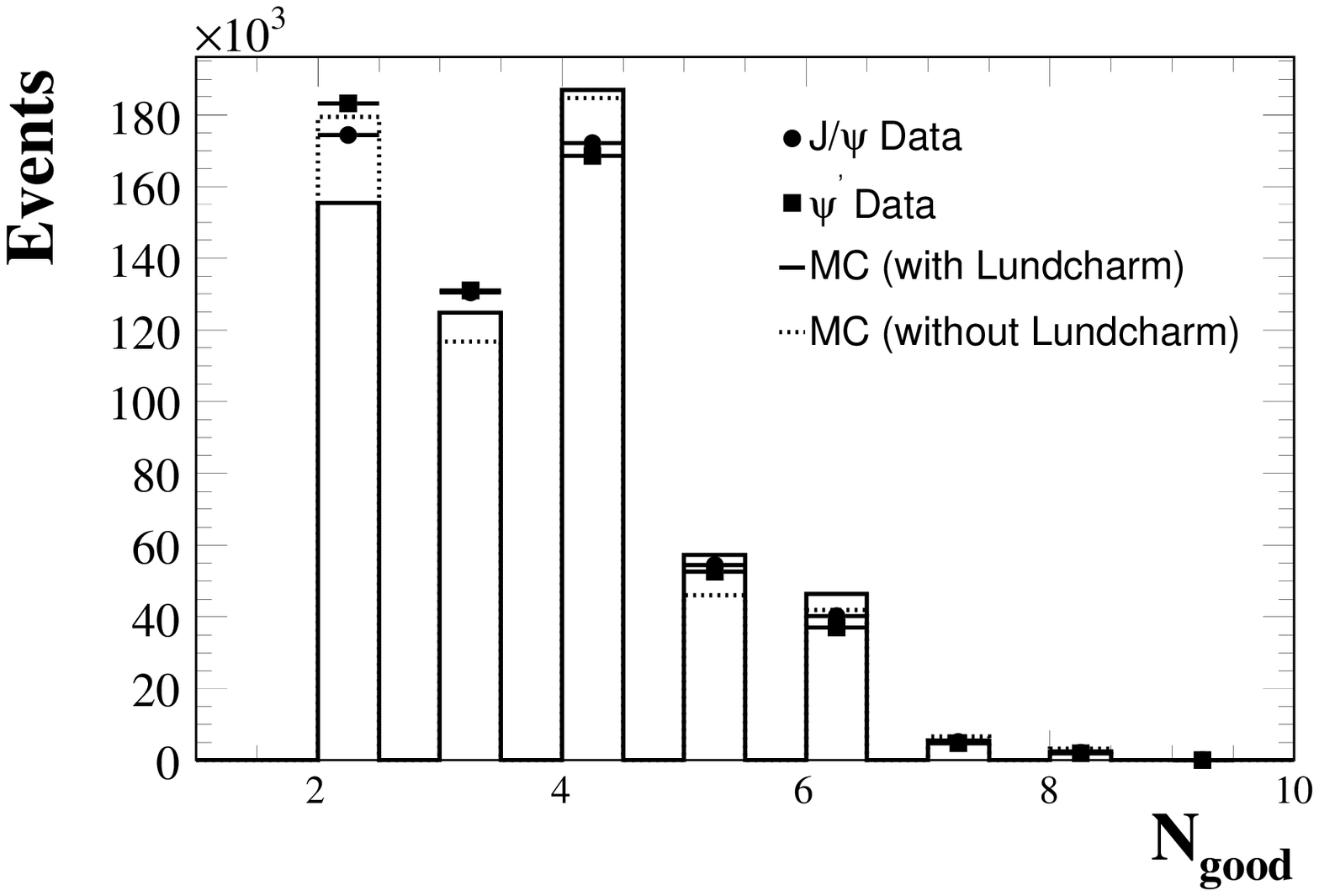}
\caption{\label{ngood} The distributions of the charged
multiplicity of $J/\psi\rightarrow inclusive$ events for $J/\psi$ data
(dots with error bars) and $\psi^{\prime}$ data (squares with error bars) and MC simulation generated with and without
the Lundcharm model (solid and dashed histograms,
respectively). }
\end{figure}

the total energy deposited in the EMC ($E_{emc}$), and
the charged multiplicity ($N_{good}$) after subtracting background
events estimated with the continuum data taken at the center-of-mass
energy of 3.08 GeV (see Section~{\ref{bkg}} for details) are shown in
Figs.~\ref{vz0} through \ref{ngood}, respectively.
Also shown are the
distributions from MC simulation, normalized to $J/\psi$ data.  The
distributions of $V_z$, $V_r$, and $\cos\theta$ of charged tracks, and
the $E_{emc}$ distribution for MC simulation
are in reasonable agreement with those from data.

For the charged multiplicity distribution shown in Fig.~\ref{ngood},
neither the MC simulation with the Lundcharm model nor the MC simulation
without the Lundcharm model agree very well with the data. However the
effect of this discrepancy between data and MC simulation on the
correction factor is very small, as described in Section~{\ref{syst}}.

\section{\label{bkg}  Background analysis}

Background events come mainly from Quantum Electro-Dynamics (QED)
processes, beam-gas interactions, and cosmic rays.  In this analysis,
all of them are estimated with the number of events selected from the
continuum data taken at the center-of-mass energy of 3.08 GeV,
normalized to the $J/\psi$ data after taking into account the
energy-dependent cross section of the QED process:
\begin{eqnarray}
\label{Nbg} N_{bg}=N_{3.08}\times
\frac{\pounds_{J/\psi}}{\pounds_{3.08}} \times
\frac{s_{3.08}}{s_{J/\psi}},
\end{eqnarray}
where $N_{bg}$ is the estimated number of background events in the
selected $J/\psi$ events; $N_{3.08}$ is the number of events selected
from the continuum data; $\pounds_{J/\psi}$ and $\pounds_{3.08}$ are
the integrated luminosities for $J/\psi$ and continuum data,
respectively; $\sqrt{s_{J/\psi}}$ and $\sqrt{s_{3.08}}$ are the
center-of-mass energies for $J/\psi$ data (3.097 GeV) and the
continuum data (3.080 GeV).

\begin{figure}[b]
\includegraphics[width=7.0cm,height=5.0cm]{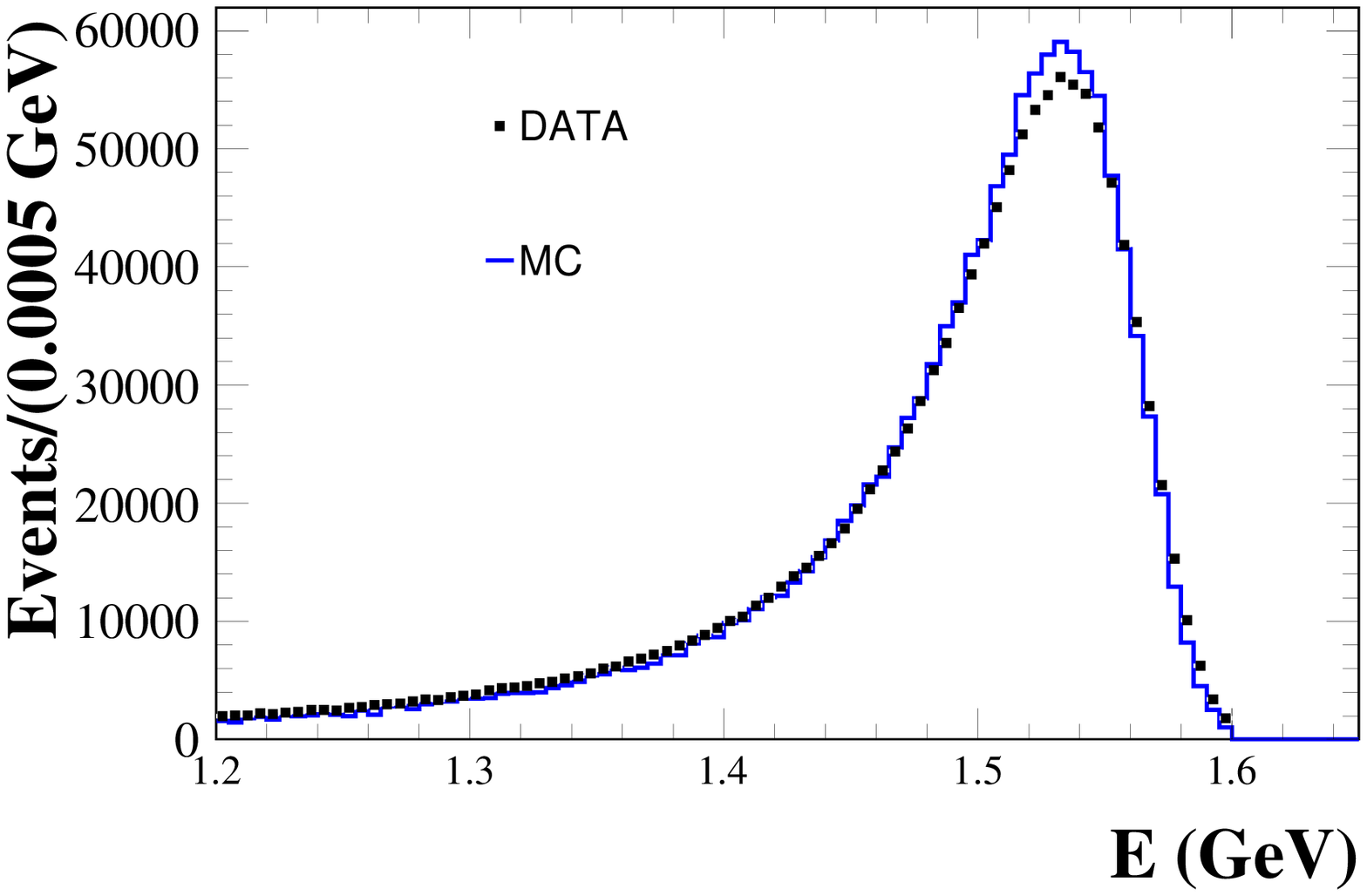}
\caption{\label{mine}The distributions of deposited energy in EMC of
photon in $e^+e^-\rightarrow \gamma\gamma$ for data (dots) and MC
simulation (histogram).}
\end{figure}

\begin{figure}[b]
\includegraphics[width=7.0cm,height=5.0cm]{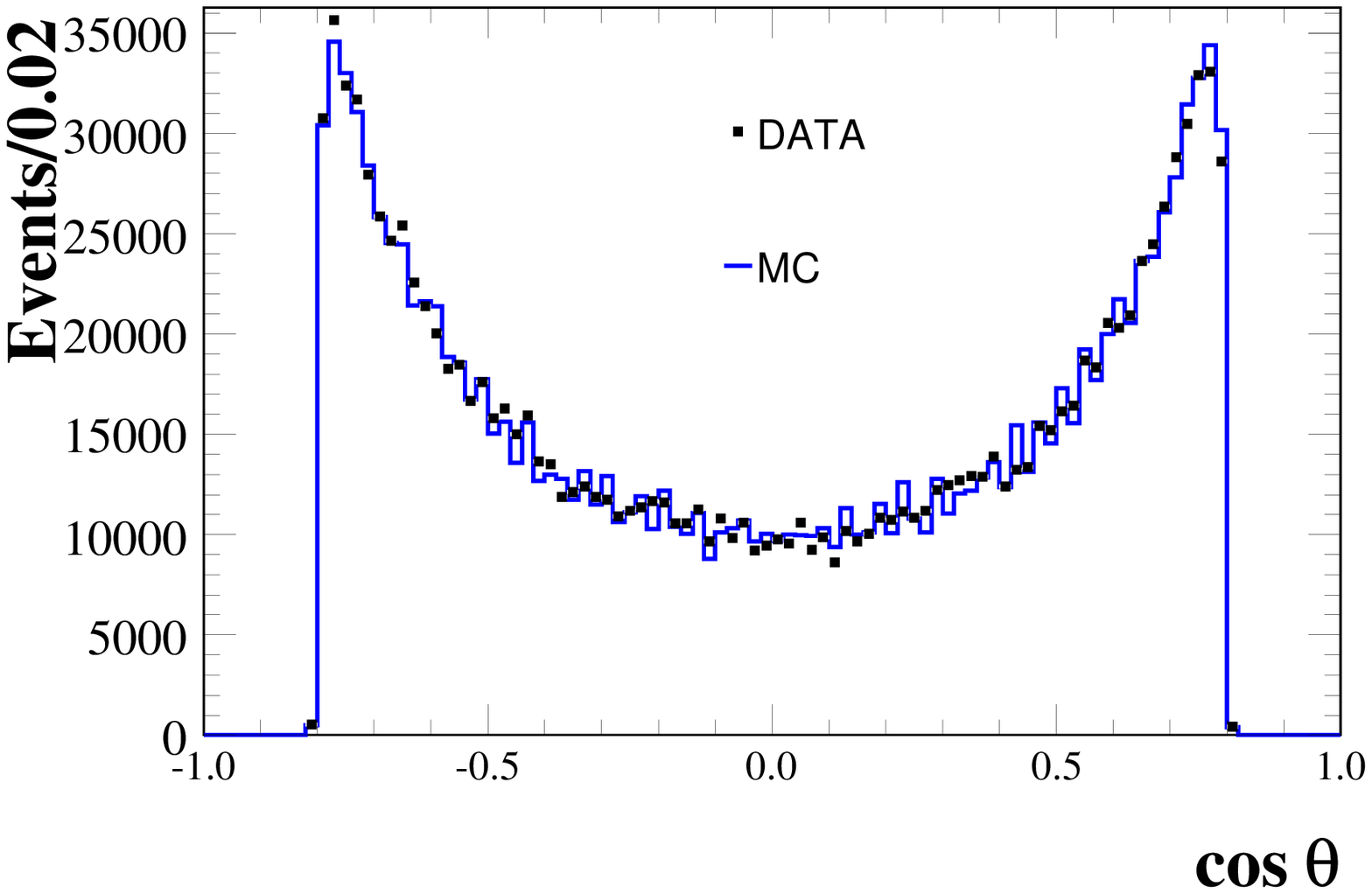}
\caption{\label{mintheta}The $\cos\theta$ distributions of
photon in $e^+e^-\rightarrow \gamma\gamma$ for data (dots) and MC
simulation (histogram).}
\end{figure}

The integrated luminosities are determined using $e^+e^-\rightarrow
\gamma \gamma$ events with the following selection criteria: there
are at least two neutral tracks with the deposited energy of the
second most energetic shower larger than 1.2 GeV and less than 1.6
GeV; and $|\cos \theta| < 0.8$, where $\theta$ is the polar angle in
the EMC. The number of signal events is determined by counting in
the signal region $|\Delta \phi| < 2.5^\circ$ and the background
estimated in the sideband region $2.5 < |\Delta \phi| < 5^\circ$,
where $\Delta \phi = |\phi_{\gamma1} - \phi_{\gamma2}| - 180^\circ $
and $\phi$ is the angle of photon in x-y plane. Figs.~\ref{mine} and
~\ref{mintheta} show the distribution of energy deposited in EMC and
$\cos{\theta}$ of photons. The integrated luminosities of $J/\psi$
data and continuum data are determined to be $79631 \pm 70~(stat.)
~nb^{-1}$ and $281 \pm 4~(stat.)~ nb^{-1}$, respectively. Here, the
statistic error is 1.5\%, and the systematic error can be cancelled
according to Eqs.~\ref{Nbg}.

With the same selection criteria for inclusive events from $J/\psi$
data, $21266\pm146$ events are selected from the continuum data.
Therefore the number of background events ($N_{bg}$) is estimated to be
$5.96 \pm 0.04$
million using Eq.~(\ref{Nbg}). The background ratio in the
selected $J/\psi \rightarrow inclusive$ events is calculated to be
3.5\% by comparing the number of background events to the
number of inclusive events selected from $J/\psi$ data.

In the above calculation, the background events from cosmic rays and
beam-gas interaction are normalized with the same procedure as QED
events. In fact, the number of cosmic rays is proportional to the data
taking time, whereas beam-gas events are related with the vacuum
status and the beam current for taking data, in addition to the data
taking time. In this analysis, the difference of the number of
background events estimated with and without considering the energy
dependence of the cross section for QED processes is taken into
account in the overall systematic uncertainty of the number of
$J/\psi$ events (see Section~{\ref{syst}} for details).

\section{Determination of the detection efficiency and correction factor}

Usually the detection efficiency is determined using a MC simulation of
$J/\psi \to inclusive$, assuming that the detector response is well
simulated. The efficiency is then the ratio between the number of
events detected and the number of events generated.  In this analysis
to avoid the uncertainty caused by any discrepancy between MC
simulation and data, the detection efficiency is determined
experimentally using 106 million $\psi^{\prime}$ events taken with the
BESIII detector. The experimental detection efficiency,
$\epsilon^{\psi^{\prime}}_{data}$, is then the number of selected
events divided by all $J/\psi\rightarrow inclusive$ events obtained
from the cascade decays of $\psi^{\prime} \rightarrow \pi^+\pi^-
J/\psi$.

\begin{figure}[b]
\includegraphics[width=7.0cm,height=5.0cm]{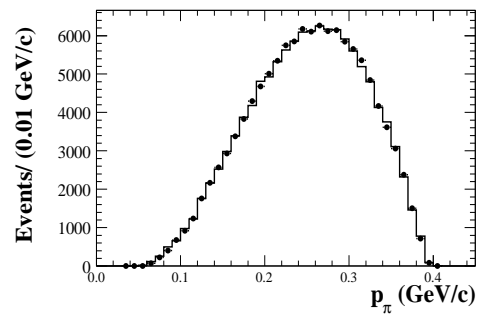}
\caption{\label{psippi} The $\pi$ momentum distributions from
$\psi^\prime$ data (dots with error bars) and MC simulation of
$\psi^{\prime} \rightarrow \pi^+\pi^- J/\psi$, $J/\psi \rightarrow \mu^+\mu^-$ (histogram). }
\end{figure}

\begin{figure}[b]
\includegraphics[width=7.0cm,height=5.0cm]{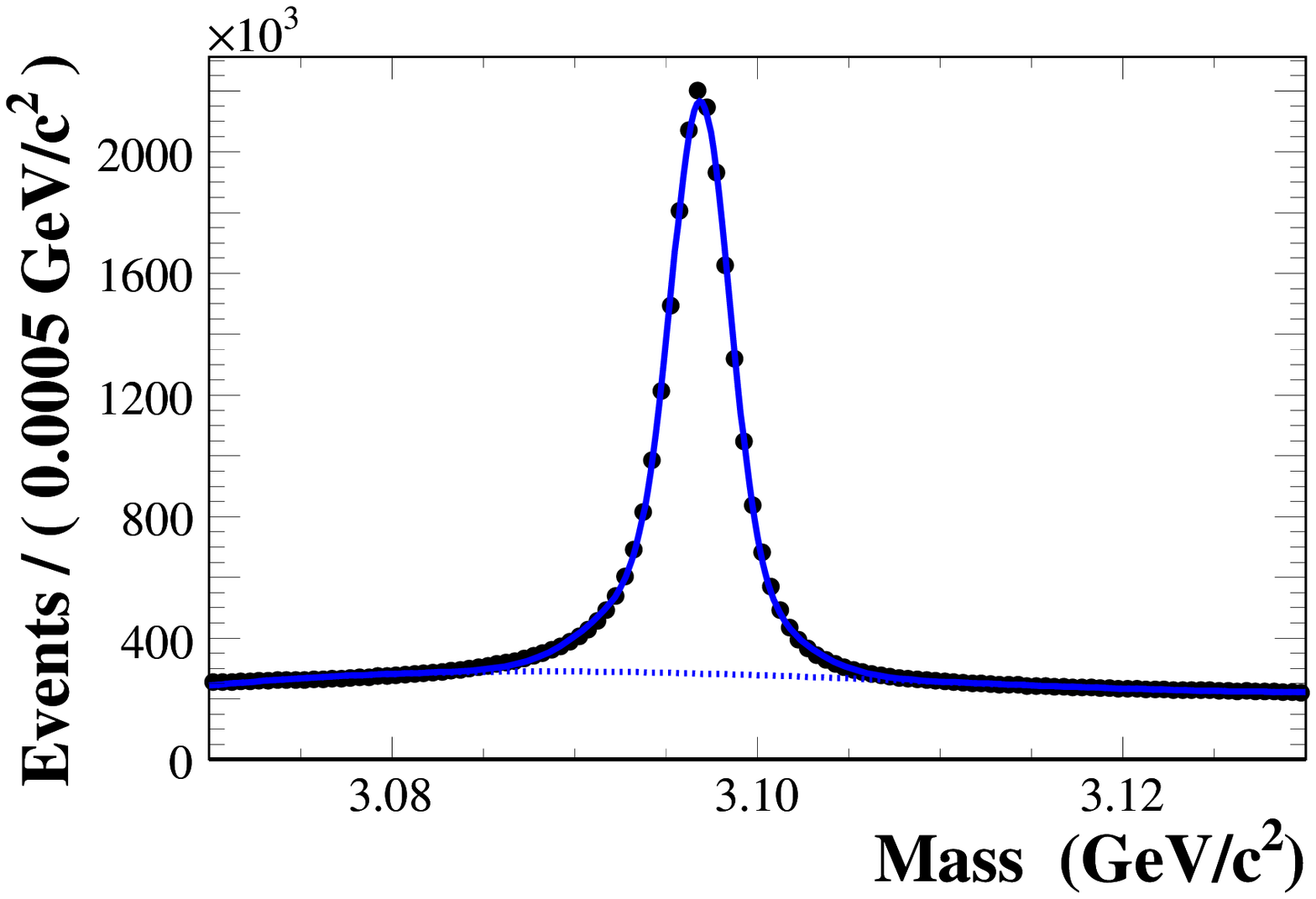}
\caption{\label{fitdat1}The invariant mass recoiling against selected
$\pi^+\pi^-$ pairs for $\psi^{\prime}$ data. A clear peak
  corresponding to  $\psi^{\prime} \rightarrow \pi^+\pi^- J/\psi$,
  $J/\psi \rightarrow inclusive$ is seen. The curves are the
results of the fit described in the text.}
\end{figure}

To select $\psi^{\prime} \rightarrow \pi^+\pi^- J/\psi$ events, there
must be at least two soft pions that are each reconstructed
successfully in the MDC within the polar angle range $|\cos\theta| <
0.93$, have $V_{r}<1$ cm and $|V_{z}|<15$ cm, and have momentum less
than 0.4 GeV/$c$. The $\pi$ momentum distributions in
Fig.~\ref{psippi} show that the MC simulation is in good agreement
with data.  There are no other requirements on the remaining charged
and neutral tracks. The invariant masses recoiling against all
possible $\pi^+$$\pi^-$ pairs are calculated and shown in
Fig.~\ref{fitdat1}. A clear peak around 3.1 GeV/$c^2$, corresponding
to the decay of $\psi^{\prime} \rightarrow \pi^+\pi^- J/\psi$,
$J/\psi\rightarrow inclusive $, is observed over a large flat
background. The number of $J/\psi\rightarrow inclusive$ events,
$N_{inc} = (19526\pm 10)\times 10^{3}$, is obtained by a fit to the
$\pi^+$$\pi^-$ recoil mass spectrum with a double-Gaussian plus a
second order Chebychev background function.

\begin{figure}[b]
\includegraphics[width=7.0cm,height=5.0cm]{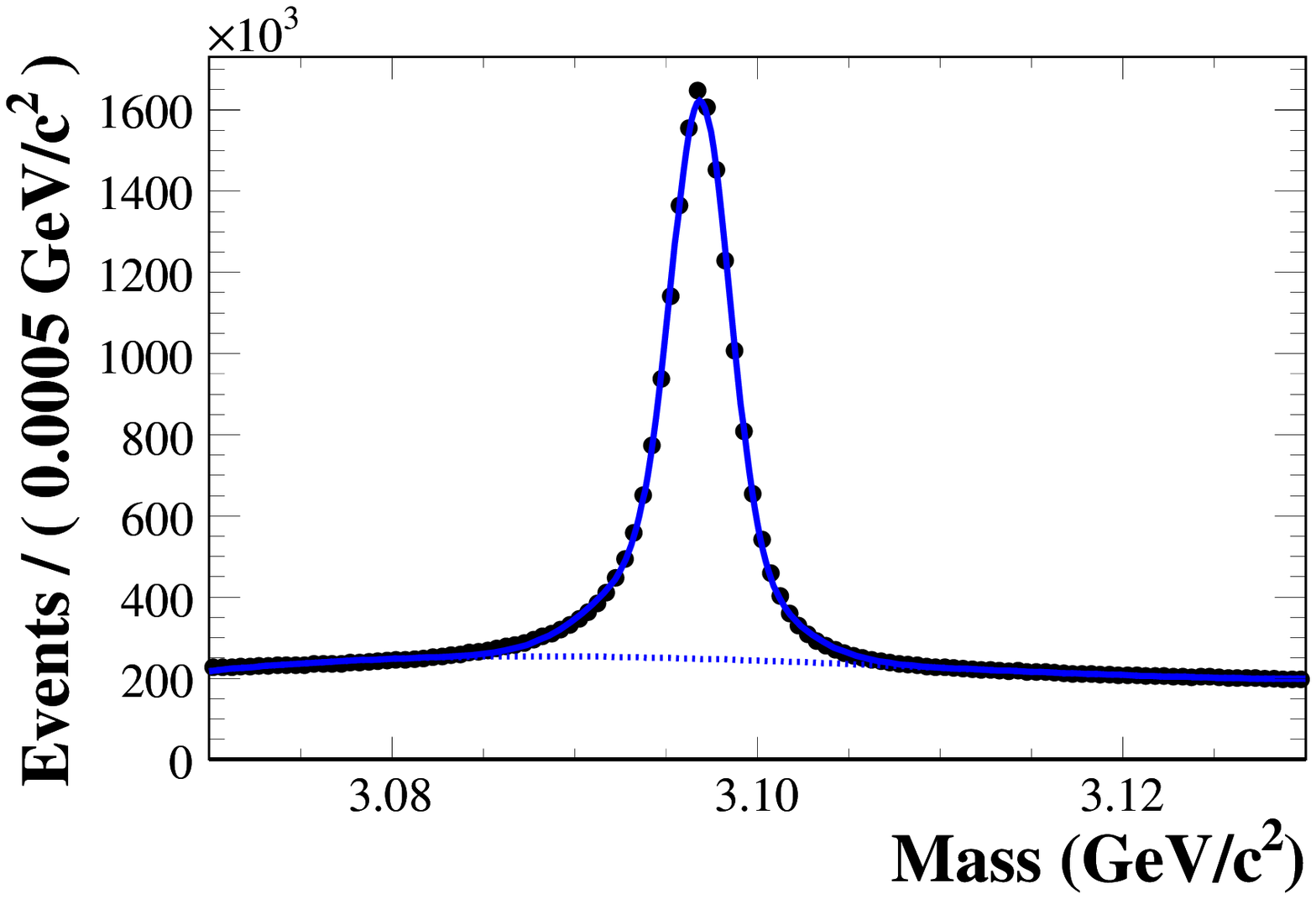}
\caption{\label{fitdat2} The invariant mass recoiling against
selected $\pi^+\pi^-$ pairs for $\psi^{\prime}$ data. Here, in
addition to selection criteria on the pion pairs, the remaining
portion of the event must satisfy the selection criteria for $J/\psi
\rightarrow inclusive$ events. The curves are the results of the fit
described in text. }
\end{figure}

To determine the number of selected $J/\psi\rightarrow inclusive$
events, in addition to the above common selection criteria for the two
soft charged pions, the remaining charged tracks and neutral tracks
must satisfy the requirements for the $J/\psi \rightarrow inclusive$
events described in Section~{\ref{bkg}}. Fig.~\ref{fitdat2} shows
the invariant mass recoiling against $\pi^+\pi^-$ for the selected
events, and the number of selected $J/\psi\rightarrow inclusive$ events,
$N_{inc}^{sel}$, is determined to be $(14432\pm 9)\times 10^{3}$
from a fit with a double-Gaussian plus a second order Chebychev
background function.  Finally the experimental detection efficiency of
$J/\psi\rightarrow inclusive$ events, $\epsilon^{\psi^\prime}_{data}$,
is determined to be $(73.91\pm0.02)$\%.

Since the $J/\psi$ decays in flight, a correction factor defined as
in Eq.~(\ref{Fcor}) is
used to correct for the kinematical effect in order to determine the detection
efficiency for direct $e^+ e^- \to J/\psi\rightarrow inclusive$
decays.
With the same procedure, including the event selection criteria and
the fit functions, the detection efficiency of
$\epsilon^{\psi^{\prime}}_{mc} = (75.87\pm 0.06)$\%, is obtained from
a MC sample of 2 million of $\psi^{\prime} \rightarrow \pi^+\pi^-
J/\psi$ events. To determine $\epsilon^{J/\psi}_{mc}$, a MC sample of 1
million events of $J/\psi\rightarrow inclusive$ was generated. With
the same selection criteria for $J/\psi \rightarrow inclusive$ events
as listed in Section~{\ref{hadsel}}, $766893 \pm 423$ events are selected, and
the corresponding detection efficiency is calculated to be
$(76.69 \pm 0.04)$\%. The correction factor $f_{cor}$ for the detection efficiency,
is then determined to be

\begin{eqnarray}
\label{Fcorr} f_{cor} = \frac {\epsilon^{J/\psi}_{mc}}
{\epsilon^{\psi^{\prime}}_{mc}}=1.0108 \pm 0.0009.
\end{eqnarray}

\section{Trigger efficiency}
The trigger efficiency of the BESIII detector has been studied using different physics
channels~\cite{trig} and was found to be very close to 100\%. Therefore, we do not repeat a similar study here,
but assume a 100\% trigger efficiency.

\section{\boldmath The number of $J/\psi$ events}
The values of different parameters used in Eq.~(\ref{Nojpsi}) are
summarized in Table~\ref{Nformula}, and the number of $J/\psi$ events
is then calculated to be $(225.30 \pm 0.02) \times 10^6$. Here the statistical
error is only from $N_{sel}$, while the statistical fluctuation of $N_{bg}$ is taken int account as
part of the systematic uncertainties (see subsection 7.4). The systematic errors from
different sources will be discussed in the next section in detail.
\begin{table}[b]
\caption{\label{Nformula}The values of different parameters used in the calculation
and the resulting number of $J/\psi$ events.}
\begin{ruledtabular}
\begin{tabular}{cc}
item & value\\ \hline
$N_{sel}$ & $(174.28 \pm 0.01) \times 10^{6}$ \\
$N_{bg}$ & $(5.96 \pm 0.04) $$\times 10^{6}$ \\
$\epsilon_{trig}$ & 1.00 \\
$\epsilon^{\psi^{\prime}}_{data}$ & $0.7391 \pm 0.0002$  \\
$\epsilon^{\psi^{\prime}}_{mc}$ & $0.7587 \pm 0.0006$ \\
$\epsilon^{J/\psi}_{mc}$ & $0.7669 \pm 0.0004  $ \\
$f_{cor}$ & $1.0108 \pm 0.0009$ \\\hline
$N_{J/\psi}$ & $(225.30 \pm 0.02) \times10^6$  \\
\end{tabular}
\end{ruledtabular}
\end{table}

\section{\label{syst}  Systematic uncertainty}
\subsection {MC model uncertainty}
The efficiency correction factor ($f_{cor}$), which is used to correct
the detection efficiency for the in-flight $J/\psi$ decay from $\psi^{\prime}$ data,
is a MC simulation dependent parameter.

  To check the MC model dependence of the correction
factor, we also determine the correction factor with MC samples
generated without the Lundcharm model. The difference of the
correction factors obtained with and without the Lundcharm model,
0.49\%, is taken as the systematic uncertainty from the MC model in
the determination of the number of $J/\psi$ events.

\subsection {Tracking efficiency}
According to tracking efficiency studies, 
the consistency of tracking efficiencies between MC simulation and data in
$J/\psi$ decays is ~1\% for each charged track, although it is a little larger at
low momentum.

In this analysis, the consistency of tracking efficiency between MC
simulation and data in $\psi^{\prime}$ decays is assumed to be the
same as that in $J/\psi$ decays. Actually there may be a difference in
the two data sets taken at different center-of-mass energies.  To estimate
the corresponding uncertainty, the tracking efficiency in the
$J/\psi$ MC sample was varied by $-0.5\%$ for the tracks with momentum
greater than $350$ MeV/$c$ and $-1.0\%$ for the tracks with momentum
less than $350$ MeV/$c$. The change of the correction factor due to
this variation leads to a change of 0.40\% in the number of $J/\psi$
events, which is taken as the systematic uncertainty due to the
tracking efficiency.

\subsection{\boldmath Fitting of $J/\psi$ peak}

From the fit of the $J/\psi$ peak we obtain the fitting errors 0.03\% and 0.08\%
in the determination of
$\epsilon^{\psi^{\prime}}_{data}$ and $\epsilon^{\psi^{\prime}}_{mc}$,
respectively, In addition, the uncertainties caused by changing the signal function,
background shape, and the fitting range in the fit of the invariant mass spectra
recoil $\pi^+\pi^-$ are also taken into account.  To estimate the
uncertainty caused by a change of the signal function, we also fit the
$J/\psi$ peak with the $J/\psi$ histogram shape, which is obtained
from the recoil mass spectrum of $\pi^+ \pi^-$ in
$\psi^{\prime}\rightarrow \pi^+ \pi^- J/\psi, ~J/\psi \rightarrow
\mu^+\mu^-$.  The change of the result is just 0.04$\%$. The
uncertainty by changing the background shape from a second order
Chebychev function to a first order one is less than 0.16$\%$. If the
fitting range is changed from [3.07, 3.13] GeV/$c^2$ to [3.08, 3.12]
GeV/$c^2$, the change is 0.32$\%$. The total systematic uncertainty
from the fitting, 0.37\%, is the sum of these errors in quadrature.

\subsection{\label{bkgerror}  Background uncertainty}

In the calculation of the number of $J/\psi$ events, the background
events from QED processes, cosmic rays and beam-gas events are
estimated by normalizing the selected continuum events by the
integrated luminosities according to Eq.~(\ref{Nbg}). Therefore the statistical error of the number
of events selected from the continuum data, 0.69\% and the uncertainties
due to the measurement of the integrated luminosities of the $J/\psi$ data and continuum
data, 1.5\%, must be taken into account in the background uncertainty.

As discussed in Section 3, normalizing cosmic rays and beam-gas events
with the energy-dependent factor for QED processes is not correct. To account
for this, the difference, 1.1\%, between the determinations of the
background normalized with and without the energy-dependent factor is
taken as a background uncertainty.

To estimate the background uncertainty from the beam-gas events, we
select samples of beam-gas events in the $J/\psi$ and continuum
data. The candidate beam-gas events must have one or two charged
tracks with the points of closest approach satisfying $|V_z|>$ 5 cm
and $|V_z|<$ 15 cm and the visible energy less than 0.5 GeV. 26844720
events are selected from the $J/\psi$ data, corresponding to 93470
events expected in the continuum data by normalizing with the
integrated luminosities.  Compared with 96230 beam-gas events directly
selected from the continuum data, the difference between them, 3\%,
is taken as a background uncertainty.

By adding all the above effects in quadrature, the total background
uncertainty is 3.6\%. Since the background ratio in $J/\psi\rightarrow
inclusive$ events is 3.5\%, the systematic uncertainty
in the number of $J/\psi$ events is 0.13\%.

\subsection {Dependence on charged multiplicity }
\begin{table}[b]
\caption{\label{difngd}The number of $J/\psi$ events and
values used in the calculation for $N_{good} \geq 2$ and
$N_{good} \geq 3$.}
\begin{ruledtabular}
\begin{tabular}{ccc}
item & $N_{good} \geq 2$ & $N_{good} \geq 3$ \\ \hline
$N_{sel}$ & 174.28$\times 10^{6}$ &119.89$\times 10^{6}$ \\
$N_{bg}$ & $5.96 \times 10^{6}$ &~$1.70 \times 10^{6}$\\
$\epsilon_{trig}$ & 1.00  & 1.00\\
$\epsilon^{\psi^{\prime}}_{data}$ & $0.7391$ &$0.5050$ \\
$\epsilon^{\psi^{\prime}}_{mc}$ & $0.7587 $  & $0.5451 $\\
$\epsilon^{J/\psi}_{mc}$ & $0.7669$ & $0.5620$ \\
$f_{cor}$ & $1.0108$ & $1.0310$\\\hline
$N_{J/\psi}$ & $225.3\times10^6$ &227.0$\times 10^{6}$ \\
\end{tabular}
\end{ruledtabular}
\end{table}

In order to reduce the number of beam-gas events in this analysis, the
selected $J/\psi\rightarrow inclusive$ events are required to have at
least two good charged tracks ($N_{good} \geq 2$). The uncertainty
from this requirement is estimated by varying the charged multiplicity
requirement from $N_{good} \geq 2$ to $N_{good} \geq 3$. For
comparison, the values obtained for the two cases are listed in
Table~\ref{difngd}. The change of the number of $J/\psi$ events,
0.76\%, is taken as the systematic uncertainty of the charged
multiplicity requirement.

\subsection {Noise mixing}

Noise in the BESIII detector has been included in the realization of
MC simulation by mixing in noise from events recorded using a random
trigger for both $J/\psi$ and $\psi^{\prime}$ data. To determine the
systematic error associated with the noise realization in MC
simulation, the $\psi^{\prime}$ MC sample is reconstructed with the
higher noise from $J/\psi$ data, and the change of the detection
efficiency correction factor, 0.4\%, is taken as a systematic
uncertainty in the determination of the number of $J/\psi$ events.

In this analysis 106 million of $\psi^\prime$ events are used to
determine the detection efficiency. However, the noise level was not
entirely stable during the period of $\psi^\prime$ data taking. To
check the effect of the changing noise level on the detection
efficiency, the $\psi^{\prime}$ data and the MC sample are divided
into three sub-samples, and the detection efficiency is determined for
each of the three samples.  The change of the detection efficiency and
the correction factor lead to a change in the number of $J/\psi$
events.  The maximum change, 0.28\%, is taken as the systematic
uncertainty associated with the changing noise levels.  The total
systematic uncertainty from the noise mixing effect is estimated to be
0.49\% by adding the individual error contributions in quadrature.

\subsection {\boldmath Estimation of $N_{J/\psi}$ with the sideband of $\bar{V}_z$}

\begin{figure}[b]
\includegraphics[width=7.0cm,height=5.0cm]{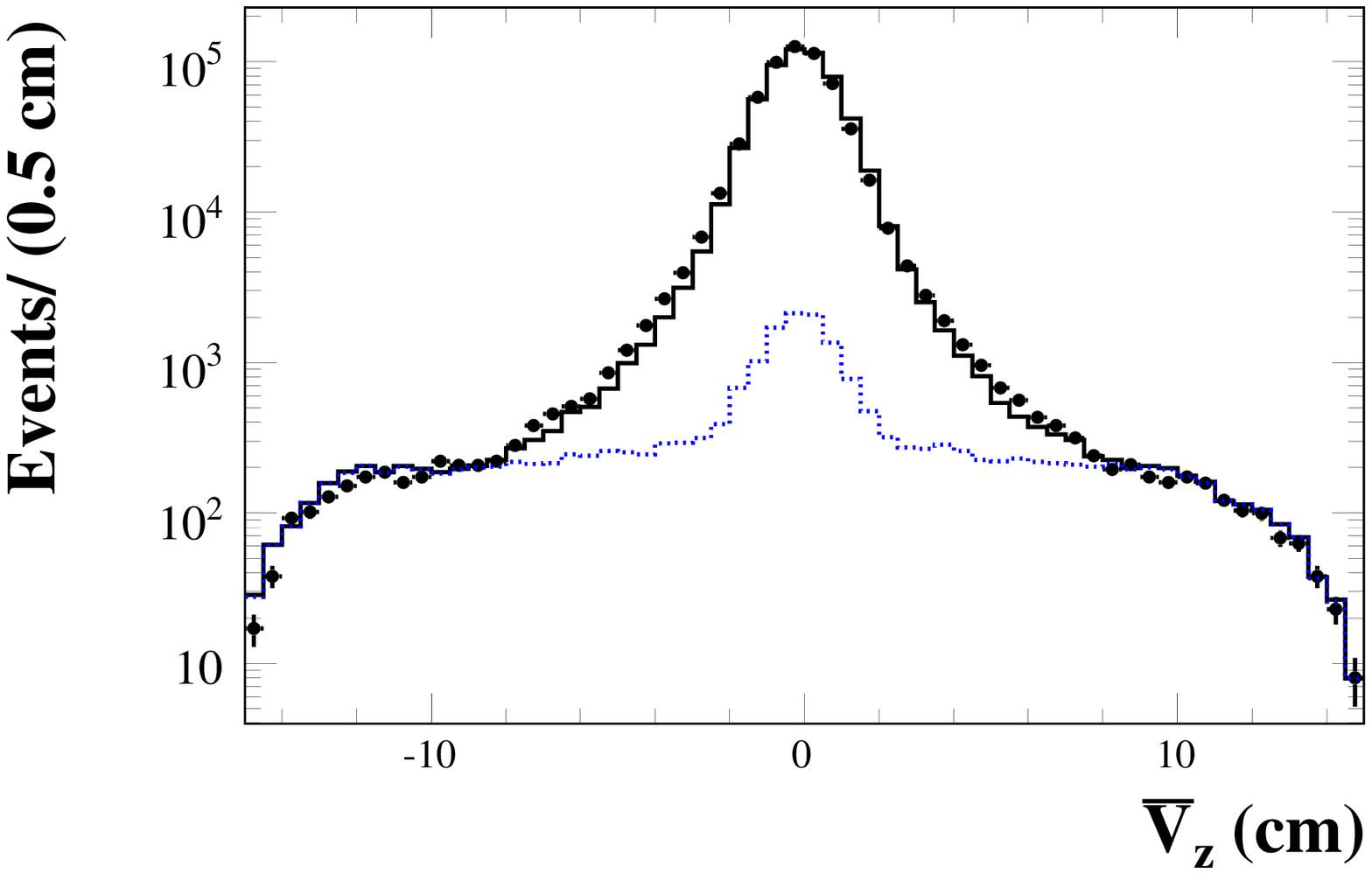}
\caption{\label{vz0bar} The distributions of the average
$z$ vertex of charged tracks, $\bar{V}_z$, for $J/\psi$ data
(dots with error bars), MC simulation of $J/\psi\rightarrow
inclusive$ plus continuum data (solid histogram) and
continuum data (dashed histogram). }
\end{figure}

The reliability of the determination of the number of $J/\psi$ events
obtained from the above method is checked by applying another method
entailing two different procedures. One difference concerns the
selection of inclusive events, which is essentially the same as in
Section 2.  except for the requirement on the track vertex position
$V_z$ along the beam direction. Here we determine the average position
$\bar{V}_z$ of the charged tracks. The signal region for inclusive
events is defined by $|\bar{V}_z|<4$ cm.  This requirement is also
applied in the determination of the detection efficiency and the
correction factor.  The $\bar{V}_z$ distribution is shown in
Fig.~\ref{vz0bar}.

The second difference is in the background estimation. The numbers of
background events from cosmic rays and beam-gas interactions are
estimated from the $\bar{V}_z$ sideband, defined by $6<|\bar{V}_z|<10$
cm. The subtraction of the sideband events from the events in the
signal region removes the cosmic ray and beam-gas events.

The sideband subtraction does not account for the QED background
events since the $V_z$ distribution is similar to that of inclusive
events from $J/\psi$ decays. However the continuum data allows us to
estimate the contribution of QED processes in the inclusive events
selected from $J/\psi$ data. The same event selection is applied to
the continuum data to select the QED events. After subtracting the
cosmic rays and beam-gas events estimated with the same sideband
method as for $J/\psi$ data, the amount of background events from the
QED processes in the selected inclusive events is estimated by
normalizing according to the integrated luminosities of the continuum
and $J/\psi$ data according to Eq. ~(\ref{Nbg}).

The same procedures have been used to determine the detection
efficiency from $\psi^\prime$ data and the correction factor with MC
samples. At last, the number of $J/\psi$ events is determined to
224.9 million. The change in the number of $J/\psi$ events with respect to
the previous method discussed in chapter 6 is 0.20\% and is
taken as a systematic uncertainty.

\subsection {\boldmath Selection efficiency uncertainty of two soft pions}

According to a MC study, the selection efficiency of soft pions,
$\epsilon_{\pi^+\pi^-}$, recoiling against $J/\psi$ in
$\psi^{\prime}\rightarrow \pi^+\pi^-J/\psi$ depends on the
multiplicity of the $J/\psi$ decay. To study its effect on the
determination of the number of $J/\psi$ events,
$\psi^{\prime}\rightarrow \pi^+\pi^- (\pi^0\pi^0) J/\psi$, $J/\psi
\rightarrow \mu^+\mu^-, 2(\pi^+\pi^-)$ events are selected from data
and inclusive MC samples, and then re-weighting factors are determined
for $J/\psi$ decaying into different multiplicities by comparing the
corresponding selection efficiency of soft pions between data and
MC. The difference between the results with and without re-weighting,
0.34\%, is taken as the uncertainty due to the selection efficiency
uncertainty of the soft pions in $\psi^{\prime}\rightarrow
\pi^+\pi^-J/\psi$.

The systematic uncertainties from different sources studied above
are listed in Table~\ref{TABSYS}. The total systematic uncertainty,
$1.24\%$, is the sum of them added in quadrature.
\begin{table}[b]
\caption{\label{TABSYS}Summary of systematic uncertainties on the
number of $J/\psi$ events.}
\begin{ruledtabular}
\begin{tabular}{cc}
Sources & Relative error (\%)\\ \hline
MC model uncertainty & 0.49 \\
Tracking efficiency & 0.40 \\
Fitting of $J/\psi$ peak & 0.37 \\
Background uncertainty & 0.13 \\
Multiplicity requirement & 0.76 \\
Noise mixing & 0.49 \\
Sideband method & 0.20 \\
$\epsilon_{\pi^+\pi^-}$ uncertainty & 0.34 \\\hline
Total & 1.24 \\
\end{tabular}
\end{ruledtabular}
\end{table}

\section{Summary}
Using $J/\psi\rightarrow inclusive$ events, the number of $J/\psi$
events collected with the BESIII detector in 2009 is determined to
be
\begin{eqnarray}
 N_{J/\psi}= (225.3\pm2.8)\times10^{6},
\end{eqnarray}
where the error is the systematic error and the statistical one is
negligible.

\begin{acknowledgments}
  The BESIII collaboration thanks the staff of BEPCII and the
computing center for their hard efforts. Supported in part by the Ministry of Science and Technology of China under Contract No. 2009CB825200; 
National Natural Science Foundation of China (NSFC) under Contracts Nos. 10625524, 10821063, 10825524, 10835001, 10935007, 11125525; 
Joint Funds of the National Natural Science Foundation of China under Contracts Nos. 11079008, 11179007; 
the Chinese Academy of Sciences (CAS) Large-Scale Scientific Facility Program; 
CAS under Contracts Nos. KJCX2-YW-N29, KJCX2-YW-N45; 100 Talents Program of CAS;Istituto Nazionale di Fisica Nucleare, Italy; U. S. Department of Energy under Contracts Nos. DE-FG02-04ER41291, DE-FG02-91ER40682, DE-FG02-94ER40823; 
U.S. National Science Foundation; University of Groningen (RuG) and the Helmholtzzentrum fuer Schwerionenforschung GmbH (GSI), Darmstadt; 
WCU Program of National Research Foundation of Korea under Contract No. R32-2008-000-10155-0
\end{acknowledgments}


\begin{thebibliography}{99}
\bibitem{bes3}Medina Ablikim et al. Nucl. Instrum. Meth. A, 2010, {\bf 614}: 345-399
\bibitem{bes2tot} Fang S S et al. HEP\& NP, 2003, {\bf 27}: 277-281
\bibitem{evtgen} Ping R G Chinese Phys. C, 2008, {\bf 32}: 599-602
\bibitem{geant4} Agostinellia S et al.,  Nucl. Instrum. Meth. A, 2003, {\bf 506}: 250-303
\bibitem{PDG}  Amsler-Gaume C et al., Phys. Lett. B, 2008, {\bf 667}: 1-5
\bibitem{LUND} Chen J C et al., Phys. Rev. D, 2000, {\bf 62}: 1-8
\bibitem{trig} Berger N et al., Chinese Physics C, 2010, {\bf 34}: 1779-1784
\end{thebibliography}
\end{document}